# Structures of the Planets Jupiter and Saturn

A Kerley Technical Services Research Report

Gerald I. Kerley

December 2004

# Structures of the Planets Jupiter and Saturn

A Kerley Technical Services Research Report

Gerald I. Kerley

December 2004

## ABSTRACT

New equations of state (EOS) for hydrogen, helium, and compounds containing heavier elements are used to construct models for the structures of the planets Jupiter and Saturn. Good agreement with the gravitational moments $J_2$ and $J_4$ is obtained with a model that uses a two-layer gas envelope, in which the inner region is denser than the outer one, together with a small, dense core. It is possible to match $J_2$ with a homogeneous envelope, but an envelope with a denser inner region is needed to match both moments. The two-layer envelope also gives good agreement with the global oscillation data for Jupiter. In Jupiter, the boundary between the inner and outer envelopes occurs at 319 GPa, with an 8% density increase. In Saturn, it occurs at 227 GPa, with a 69% density increase. The differences between the two planets show that the need for a density increase is not due to EOS errors. It is also shown that helium enrichment cannot be the cause of the density increase. The phenomenon can be explained as the result of enrichment of heavy elements from planetismals that were dispersed and mixed with the gases during planet formation. This conclusion is consistent with planetary formation models like that of Pollack, et al.

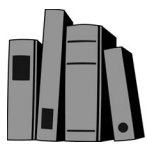

**K**erley
**T**echnical
**S**ervices
P.O. Box 709
Appomattox, VA 24522-0709

Tel: (434) 352-8400 ■ Fax: (434) 352-4973 ■ E-mail: gkerley@compuserve.com

# CONTENTS

# FIGURES

# TABLES

# 1. INTRODUCTION

## 1.1 Background

Speculation about the planets Jupiter and Saturn, what they are and how they came to be, has existed since ancient times. The invention of telescopes, which allowed observation of their beauty, and the discovery of the laws governing planetary motion, greatly aroused interest in these planets. Subsequent advances in observational methods, physical theories, and numerical computation not only intensified this curiosity but also allowed the development of quantitative planetary-structure models.[1] In recent times, the Pioneer, Voyager, Galileo, and Cassini space missions, along with the impact of the Shoemaker-Levy 9 comet fragments into Jupiter, have stimulated the greatest level of interest thus far. An understanding of the structure of the Jovian planets could also help to answer questions about their formation and evolution, as well as that of planets recently discovered outside our solar system.

My own interest in these problems originally developed from a desire to see if the planets could shed any light on the validity of equation of state (EOS) models for hydrogen and deuterium. While recent experimental and theoretical work has greatly improved understanding of the EOS for those materials, pressure-temperature states in the deep interiors of the Jovian planets still cannot be accessed in terrestrial laboratories at present. Of course, the planetary structure problem is much too complicated to permit definitive answers to questions about the validity of EOS models. However, the present study demonstrates that my new $H_2$ EOS model [2] is at least consistent with the observational data currently available. It also offers new insights into the structures of Jupiter and Saturn that should be of interest to the astrophysical community.

## 1.2 Elements of a Planetary Model

Standard models for Jovian planets involve four main features [3]-[7]:

- All points in the planet's interior are assumed to lie on a single pressure-density curve. This curve is usually taken to be an adiabat (isentrope) passing through a particular pressure-temperature state at the surface. Hence I will refer to this curve as the "planetary adiabat," although the entropy can be discontinuous at points where the chemical composition changes. The surface state is determined from observational data.

- The chemical composition (element abundances) along the adiabat is essentially a model parameter. It is relatively easy to show that the Jovian planets must be composed mostly of hydrogen and have dense cores.

---

1. See Chapter 10 of of Ref. [1] for an overview of work from 1923 to 1983.

Some inferences can be obtained from surface data and hypotheses about planetary origins. In the end, however, the abundances must be chosen by fitting the model to the observed mass, radius, period, and gravitational moments.

- Once the element abundances have been specified, the planetary adiabat is calculated using a particular EOS model. This model involves EOS for the individual chemical elements and compounds, together with a mixture algorithm. The EOS model has also been treated as a fitting parameter in some previous work [4].

- Given the planetary adiabat, the pressure and density are determined as functions of radius and polar angle by requiring the mass distribution to give hydrostatic equilibrium in the planet's own gravitational field. The dependence on polar angle, which arises from the rotational motion, is usually handled by a perturbation method called the theory of figures, as described by Zharkov and Trubitsyn (ZT) [3].

All of the above features have uncertainties that affect the accuracy of the model; errors in one part can compensate for errors in another.

The ZT theory of figures is most certainly the most reliable part of the model. The two main approximations in the theory are 1) truncation of the perturbation series, and 2) neglect of differential rotation. The third-order treatment of the rotational perturbation [3], is used in this work.[1] This approximation is quite accurate for the first two (even) gravitational moments, $J_2$, and $J_4$. The accuracy is less for $J_6$ but should still be within current experimental errors. I have also ignored the effects of differential rotation, which are currently thought to be small.[2]

The temperature along the planetary adiabat is a possible source of uncertainty. The assumption of an adiabatic temperature profile is open to question. Guillot, et al. [9], showed that deviations from the adiabatic assumption, due to radiative effects, could occur in both Jupiter and Saturn. However, I will consider this possibility in Sec. 5. and show that it is probably not a significant cause for concern in the context of the present study. I will also show that uncertainties in the surface temperature have only a small effect on the model predictions.

Since there is no way to measure the elemental abundances in the interiors of the planets, the chemical composition along the planetary adiabat is clearly a *major* source of uncertainty in the model. The importance of the chemical composition was demonstrated in a recent study by Saumon and Guillot [7]. They showed that two parameters—the masses of heavy elements in the core and in the gas envelope—could be varied to fit the observed values of $J_2$ and $J_4$ for Jupiter and Saturn. But very different values were obtained when using different EOS for hydrogen and helium. Hence uncertainties in the chemical composition have essentially the same effect as uncertainties in the EOS.

---

1. One of the equations in Ref. [3] contains a typographical error. The correct equation, given in Sec. 3., was used in this work.

2. Differential rotation is a variation of the rotational period with latitude and depth. Hubbard has argued that this phenomenon can be treated by making small adjustments to the gravitational moments obtained in a solid rotation model [6][8]. Those adjustments would not affect the central conclusions of this report.

In establishing the chemical composition, one must lean heavily on measurements taken at the surface. Unfortunately, detailed data are presently available only for the surface of Jupiter, obtained from the Galileo probe. (Even in that case, the oxygen abundance data apparently do not apply to the entire surface.) Moreover, there is no justification for assuming that the surface composition is applicable throughout the entire gas envelope. In fact, conventional wisdom among planetary modelers is that the gas envelope is *not* homogeneous in either planet.

In previous work, the EOS of hydrogen and helium have been regarded as the most important sources of uncertainty in models of Jupiter and Saturn. In the past, distrust of the EOS models could be attributed both to a lack of experimental data and to wide variations in the predictions of different models. In recent years, however, advances in both experiment and theory have led to a much improved understanding of hydrogen and helium at high pressures and/or temperatures. There is good reason to believe that, for the purposes of planetary modeling, EOS uncertainties are now less significant than uncertainties in chemical composition.

## 1.3   Scope of this Investigation

The hydrogen and helium EOS tables used in this work, developed quite recently, give better agreement with experimental data and *ab initio* numerical calculations than those used in previous planetary modeling studies. The present study also employs a new and more realistic treatment of the heavy elements and compounds present in the planetary gas envelopes, as well as in the central core.

Given this EOS model, the present study seeks to determine the elemental abundances that give the best agreement with the measured values of the gravitational moments $J_2$, $J_4$, and $J_6$. The global oscillation data for Jupiter are also considered. Very good results for both planets are obtained by using a two-layer gas envelope in which the inner region has a higher density than the outer one. The density increase is attributed to the presence of heavy elements from planetismals that were dispersed and mixed with the gas envelope during formation of the planets. Helium enrichment, while not completely ruled out, cannot account for this effect.

It is also found that increasing the abundance of heavy elements in the inner region of the gas envelope reduces the size of the central core that is required to match the known planetary masses. In fact, the best agreement is obtained by al-lowing the central core to be as small as possible. Both planets are assumed to have central cores of about 3 $M_E$ ($M_E$ = mass of the earth, $5.9742 \times 10^{27}$ g [10]). However, the *total* mass of heavy elements, in the central core and inner region of the gas envelope, is roughly 30 $M_E$ for both planets. This result is consistent with the predictions of planetary origins using accretion models [11][12].

The remainder of this report is organized as follows.

- Section 2 gives an overview of the EOS for the individual chemical species, the mixture/chemical equilibrium model, and EOS for the core. Additional details are given in Appendix A.

- Section 3 reviews the theory of figures and discusses the numerical methods used in this work.

- Section 4 defines "baseline" models, consisting of a homogeneous gas envelope and a central core, for the two planets. The elemental abundances obtained from surface measurements are shown to give poor agreement with the measured gravitational moments.

- Section 5 considers several attempts to improve the predictions while retaining the assumption of a homogeneous gas envelope. It is possible to match $J_2$ with these models, but none of them can match both $J_2$ and $J_4$.

- Section 6 discusses the issue of hydrogen-helium demixing. It is shown that the adiabats for both planets lie above the region where phase separation is expected to occur and that demixing is not a likely mechanism for helium enrichment in the planetary interiors.

- Section 7 discusses the two-layer envelope model from several points of view—a density-scaling argument, calculations based on specific assumptions about the elemental abundances in the inner region, and the effect of a transition zone between the two regions. It is shown that the two-layer model gives good agreement with the gravitational moments for both planets and also the global oscillation data for Jupiter.

- Conclusions are summarized in Section 8.

Tables of the planetary adiabats and the planetary structures are given on the CD enclosed with this report. The CD also includes electronic copies of this report and the reports on the hydrogen and helium EOS.

## 2. THE EOS MODEL

### 2.1 Hydrogen EOS

The EOS for molecular and atomic hydrogen, $H_2$ and H, were taken from Ref. [2]. However, the $H_2$ EOS was modified to include corrections to the rotational motion, as discussed below.

The hydrogen EOS of Ref. [2], constructed using the PANDA code [13], employs a multiphase, multicomponent, chemical equilibrium model that treats dissociation, ionization, and the insulator-metal transition. The main features of the model are as follows.

- Separate EOS were constructed for the molecular solid, the monatomic solid, the molecular fluid, and the monatomic fluid. They were combined into a single table using the principles of phase transitions and chemical equilibrium.

- The EOS for the molecular solid includes contributions from the zero-Kelvin isotherm, lattice vibrations, and internal vibration and rotation.

- The EOS for the monatomic solid includes contributions from the zero-Kelvin isotherm, lattice vibrations, and thermal electronic excitation and ionization. The insulator-metal transition is included in the thermal electronic term.

- The EOS for the molecular and atomic fluids were constructed from a variational theory of liquids called the CRIS model [14][15]. The fluid EOS also include the same vibrational-rotational and thermal electronic terms as in the solid EOS.

- The EOS for the molecular and atomic fluids were combined into a single EOS for the fluid phase using the ideal mixing approximation and the assumption of chemical equilibrium. The model predicts a transition from a molecular fluid to a metallic atomic fluid at high pressures but does not predict phase separation between the two constituents in the fluid phase.

The above model was used to construct an EOS for deuterium as well as hydrogen. Reference [2] gives an extensive comparison of the models with experimental data for both isotopes. The models give very good agreement with melting and vaporization data, low-temperature static compression and sound speed measurements, principal Hugoniots, reshock and reverberation experiments, and even conductivity experiments. The reshock and reverberation experiments are of special significance in the context of the present study, because they generated pressures as high as 400 GPa, with temperatures close to those on the Jupiter adiabat. The fact that the EOS model, which was not fitted to those data, is in good agreement with the measurements, lends credence to its validity for use in planetary modeling.

Reference [2] also showed that the EOS model gave very good agreement with numerical results generated using the Path Integral Monte Carlo (PIMC) method, the best numerical technique presently available.[1] After the report was written, additional PIMC calculations, of Hugoniots for precompressed liquid deuterium, were discovered [17]. The EOS model also gives good agreement with those calculations, which also reach pressures of about 400 GPa, at temperatures close to those on the planetary adiabats. Once again, the comparison lends credence to the validity of the EOS used in this work.

However, the present work did uncover a problem with the EOS model of Ref. [2] at low temperatures. When used to compute the Jupiter adiabat, it did not give satisfactory agreement with the temperature vs. pressure in Jupiter's atmosphere, as determined by the Galileo probe from 1 to 22 bars. (See Sec. 4.1.) Upon examination, it was discovered that the discrepancy arose from the neglect of nuclear spin effects in the contributions from rotational motion. This problem was resolved by using an exact calculation of the rotational partition function, for an equilibrium mixture of ortho-$H_2$ and para-$H_2$, to compute corrections to the $H_2$ EOS of Ref. [2]. Details are discussed in Appendix A.

The rotational correction terms have no effect on the comparisons with experiment given in Ref. [2] because they do not contribute to the pressure and make only small contributions to the energy and entropy at temperatures above 500K. However, they are significant for the present study, because they determine the temperature on the planetary adiabat. Without the corrections, the model would predict too low a temperature, and thus too high a density, at a given pressure on the adiabat. The higher temperatures are also relevant to the issue of hydrogen-helium demixing, discussed in Sec. 6.

The present work uses only the EOS tables for the molecular and atomic fluids, together with EOS tables for the other elements and compounds and the PANDA mixture model. The solid phases are not needed because the adiabats for Jupiter and Saturn both lie entirely within the fluid regime.

---

1. Another popular numerical method combines density functional theory with molecular dynamics (DFT/MD). This technique, unlike PIMC, involves approximations in the underlying theory as well as in the numerical methods. Hence it generates systematic errors and should not be regarded as an exact method. For example, current density functionals underestimate the ionization energy of the hydrogen atom by a factor of 2 [16]. As a result, DFT/MD calculations of the EOS can be expected to overestimate the amount of temperature and/or pressure ionization. DFT/MD calculations of the deuterium Hugoniot do not give satisfactory agreement with either the experimental data or the EOS model of Ref. [2]. Overestimation of ionization effects is one likely cause of the discrepancies.

## 2.2  Helium EOS

The helium EOS, taken from Ref. [18], was constructed using the variational theory of fluids (CRIS model) and including contributions from thermal electronic excitation (PANDA ionization equilibrium model). The model was similar to that used for atomic hydrogen, except that the interatomic forces are more similar to those for $H_2$ than H. In fact, the attractive forces between He atoms are so weak that the critical point of liquid He is only 5.2K, compared to 33.2K for $H_2$.

A model was also constructed for solid He but was not used in the present work because the planetary adiabats lie in the fluid regime.

Further details are given in Ref. [18].

## 2.3  Mixture Model for Envelope

The envelope region of the Jovian planets is a mixture of hydrogen, helium, and compounds containing heavier elements, the most abundant being carbon, oxygen, nitrogen, and sulfur. These elements can exist in several chemical forms. At low pressures and temperatures, the predominant molecular species are $H_2$, $H_2O$, $CH_4$, $NH_3$, and $H_2S$. At high pressures and/or high temperatures, these molecules dissociate, giving an atomic mixture. (Helium forms no compounds and exists only in the atomic form.) Ionization also occurs at sufficiently high pressures and temperatures.

The EOS for the envelope region was calculated using the mixture-chemical equilibrium model in the PANDA code [13]. The model used in this work, an updated version of the one given in the original PANDA manual, is discussed in Sec. 9 of Ref. [2]. Only the main points are summarized here.

The PANDA model constructs a mixture EOS from separate EOS tables for each chemical species to be allowed in the chemical equilibrium calculation. In the present work, EOS tables for $H_2$, H, and He were taken from Refs. [2] and [18], as described above. The EOS tables for $H_2O$, $CH_4$, $NH_3$, N, and O were taken from Refs. [19]-[21], and the EOS table for atomic carbon was taken from Ref. [22].[1] The EOS tables for $H_2S$ and S were constructed from the tables for $H_2O$ and O by mo-lecular weight scaling. This approximation is adequate for the present work, since the sulfur compounds make a very small contribution to the EOS.

---

1. Some improvements have been made to the original models for $H_2O$, $CH_4$, $NH_3$, N, and O. These changes will be discussed elsewhere. Sensitivity studies showed that these changes have a negligible effect on the calculations discussed in this report. Only the liquid phase of carbon was used in these calculations. Sensitivity studies showed that the solid phases are not formed under conditions relevant to the Jovian planets. Sensitivity studies also showed that other compounds, e.g., $N_2$, $O_2$, $CO_2$, NO, and HCOOH, were not present in sufficient quantities to have a significant effect on the EOS.

Contributions from electronic excitation and ionization are included in the individual EOS tables for the atomic species. Hence the model accounts for both pressure ionization and thermal ionization.

PANDA uses the ideal mixing approximation to construct an EOS for the mixture from the EOS tables for the individual components. The formation of an ideal mixture (also called an ideal solution) can be regarded as a two-step process. First form a heterogenous mixture in which all components have equal pressures and temperatures. Second, allow the components to mix homogeneously, at constant temperature and pressure. In ideal mixing, the second step involves no change in volume or energy, while the entropy change corresponds to complete randomness. Despite its simplicity, this model has been found to give good agreement with precise Monte Carlo calculations of mixtures, using hard sphere, soft sphere, and exponential-6 potentials [21][23].

The chemical composition of the mixture is determined from the principle of chemical equilibrium—at each density and temperature, the Helmholtz free energy is minimized with respect to the mole fractions, subject to constraints that follow from the chemical formulas of the species. In setting up a problem, the user enters the molecular formula for each mixture component, along with any "initial" composition that is consistent with the overall elemental composition of the system. PANDA then determines the chemical constraints from the molecular formulas and the initial composition and reduces the constraints to a linearly independent set of equations.

The PANDA model also offers a parameter (PTYPE) that can be used to treat immiscibility by modifying the entropy of mixing contribution for each component (see Eq. 56 of Ref. [2]). For PTYPE=1.0 (the default), a component is completely miscible with the other components. For PTYPE=0.0, it forms a separate phase. Intermediate values correspond to partial miscibility. This parameter affects the EOS in that it alters the equilibrium composition at a given pressure and temperature—lowering the value of PTYPE reduces the entropy, and thus the stability, of a given chemical component.

In this work, I have used PTYPE=0.75 for atomic hydrogen, to be consistent with Ref. [2]. Changing to PTYPE=1.0 affects the gravitational moments by 1.0-2.0%; this effect is small, when compared with the effect of varying the element abundances, and so is insignificant for the purposes of this report.[1]

--------

1. I have also used PTYPE=0.05 for carbon, to be consistent with other work on CHON mixtures. Sensitivity studies showed that the value of this parameter has a negligible effect on the calculations.

Figure 1 shows the concentrations of all chemical species (except $H_2S$ and S) as functions of pressure on the Jupiter and Saturn adiabats, for the baseline models discussed in Sec. 4. For purposes of illustration, the mixture calculations have been continued on beyond the pressures at the core boundary, denoted by vertical dotted lines (4700 GPa for Jupiter, 860.1 GPa for Saturn). Molecular hydrogen dissociates over a wide range of pressures, 150-1500 GPa, the molecular and atomic forms having equal concentrations at ~400 GPa. Hence $H_2$ is completely dissociated at the core boundary in Jupiter but only partially dissociated in the case of Saturn. $CH_4$ dissociates at ~250 GPa, $NH_3$ at ~3000 GPa, and $H_2O$ at ~13,000 GPa. Hence $CH_4$ is dissociated in both planets, $NH_3$ is dissociated in Jupiter but not in Saturn, and $H_2O$ is not dissociated in either planet.

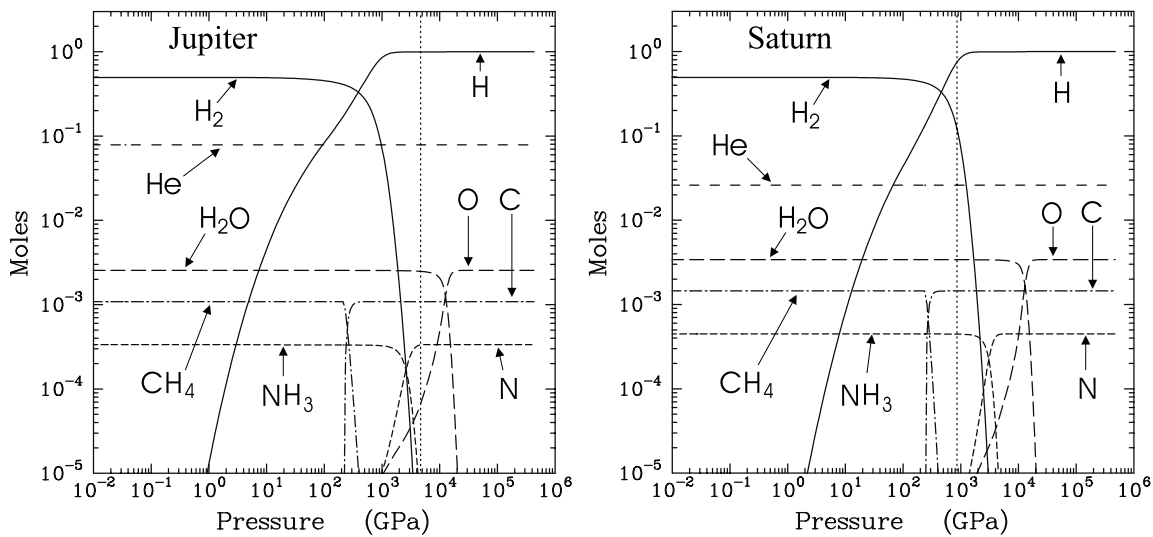

**Fig. 1. Concentrations vs. pressure in planetary envelopes, using baseline models of Jupiter (left) and Saturn (right). Core pressures are denoted by vertical dotted lines.**

## 2.4 Model for Core

Calculations for the Jovian planets typically require the existence of a dense core to account for the total mass. A core is also to be expected from currently favored theories of planetary formation [11][12]. It is thought that a protoplanet was initially formed by accretion of solid planetismals from the solar nebula. This protoplanet, consisting primarily of ice and rock, then became the core that accreted gases, eventually forming the envelope.

The above scenario would lead to a planetary core consisting of water and silicates, along with some carbon and nitrogen. (Pollack, et al. [12], proposed a composition similar to that of comet Halley.) The exact composition, of course, cannot be measured.

To test the importance of the core composition in modeling the planets, I compared calculations using three different core materials—water, silica ($SiO_2$), and iron—using EOS tables appropriate for the high core pressures and temperatures.[1] The gravitational moments and equidistance differed only slightly for the three cases, despite differences in the core masses and radii.[2]

These results show that the core composition and EOS are much smaller sources of uncertainty in modeling the Jovian planets than the parameters that define the gas envelope. Except where indicated, the results presented in this report were computed using a silica core.

## 2.5  Calculation of Planetary Adiabats

A planetary adiabat consists of two or more segments or regions. In the simplest case—a homogeneous gas envelope and dense core—the PANDA isentrope option was used to compute and tabulate separate adiabats for the two regions. These adiabats were joined, using the envelope at the lower pressures and the core at the higher pressures, matching the pressure and temperature at the core boundary.

The planetary equilibrium calculation used in this work employs an iterative procedure in which the pressure at the core boundary is adjusted at each iteration, to match the total mass. (See Sec. 3.) Since the two adiabats are computed prior to the calculation, this procedure can give a slight temperature mismatch at any given iteration. Fortunately, the core adiabat has a weak dependence on temperature at the high core pressures, so this effect presents no problems in practice.

In the two-layer envelope model, discussed in Sec. 7., separate adiabats were computed for the two regions, using different element abundances, then joined by matching the pressure and temperature at the boundary between the inner and outer regions. Joining to the core region was done just as for the homogeneous envelope.

---

1.  Water and silica can dissociate at the high pressures and temperatures reached in the planetary cores, and this effect must be included in the EOS table. The EOS of $H_2O$ was taken from Ref. [24], that for $SiO_2$ from Ref. [25], and that for Fe from Ref. [26]. The corresponding EOS tables in the Sandia/CTH version of the Sesame library [27] are: $H_2O$—#7150, $SiO_2$—#7360, Fe—#2150. Note that EOS tables 7360 and 2150 are not available in the Los Alamos version of the Sesame library [28].

2.  For the baseline Jupiter model discussed in Sec. 4., the results were: mass fractions—0.092 ($H_2O$), 0.078 ($SiO_2$), 0.066 (Fe); core radius/outer radius ratios—0.21 ($H_2O$), 0.17 ($SiO_2$), 0.12 (Fe). The results for the baseline Saturn model were: mass fractions—0.31 ($H_2O$), 0.28 ($SiO_2$), 0.25 (Fe); core radius/outer radius ratios—0.27 ($H_2O$), 0.24 ($SiO_2$), 0.16 (Fe). The differences are less for the other planetary models, which have smaller cores.

A discontinuous change in density at the transition pressure leaves the sound speed at the transition undefined. In order to avoid this difficulty, the transition was spread out over a 0.2% range of pressure, and an interpolation formula was used in the transition region. The sound speed in the transition region was then computed from the square root of the numerical derivative of the pressure with respect to density. Hence the sound speed in the transition region is lower than in the adjacent regions of the envelope. This procedure gives a slightly lower result for the equidistance. (See Sec. 3.)

In the calculation using a radiative region, the envelope consists of three segments, as discussed in Sec. 5.3. Here again, the three segments were joined by matching the pressure and temperature at the two boundaries.

# 3. THE THEORY OF FIGURES

The present investigation employs the third-order theory of figures, taken from the 1978 book by Zharkov and Trubitsyn (ZT) [3]. This section reviews the equations given in ZT and corrects a typographical error that appears in one of them. It also discusses the numerical procedures that were used to solve those equations.

## 3.1 Basic Equations

The following discussion gives the most important equations from ZT. The equation numbers used in their book are also given for reference.

The total potential at a position $r$ within a planet is the sum of gravitational and centrifugal terms,

$$U(r) = V(r) + Q(r).$$
[ZT24.1]    (1)

The gravitational potential is

$$V(r) = G \int [\rho(r')/|r - r'|] d^3 r',$$
[ZT24.2]    (2)

where $G$ is the gravitational constant ($6.67259 \times 10^{-11}$ m$^3$/kg/s$^2$ [10]), $\rho(r)$ is the density at position $r$, and the integral is taken over the entire planet. If the rotation rate is the same at all points within the planet, the centrifugal potential is

$$Q(r) = \tfrac{1}{2}\omega^2 r^2 \sin^2\theta,$$
[ZT24.2]    (3)

where $\theta$ is the angle from the axis of rotation. The angular velocity $\omega$ is related to the period of rotation $d$ by $\omega = 2\pi/d$.

The structure of the planet, i.e., the density function $\rho(r)$, is to be calculated from the equation of hydrostatic equilibrium,

$$\rho^{-1}\nabla p = \nabla(V + Q) = \nabla U,$$
[ZT24.4]    (4)

where the relationship between pressure $p$ and density is given by the planetary adiabat. Hydrostatic equilibrium is valid if the planet is in a gas-liquid state (does not support shear stresses) and is evolving (contracting, radiating, etc.) very slowly in time. These conditions are satisfied for Jupiter and Saturn.

In the absence of rotation ($\omega = 0$), the planet is spherically symmetric, and the density is a function only of the radius $r$. Rotation deforms the planet into a spher-oid that is flatter at the poles. In that case, the density depends on two variables, $r$ and $\theta$. (A second angle is not needed because the planetary structure does not depend on longitude.)

The theory of figures expands the rotational effects in powers of small parameters $q$ and $m$,

$$q = \omega^2 a_1^3 / GM_1, \qquad m = \omega^2 s_1^3 / GM_1 = q(s_1/a_1)^3 \qquad \text{[ZT24.3]} \quad (5)$$

where $M_1$ is the planetary mass, $a_1$ is the equatorial radius, and $s_1$ is the average radius (defined below). (ZT use the subscript 1 to specify values of the mass and radius at the planetary surface. Unsubscripted symbols, as used below, indicate variables at any point within the planet.)

The theory seeks equations for the equipotential surfaces, i.e., spheroidal surfaces on which the total potential $U(r, \theta)$ is a constant. The solutions are assumed to have the form

$$r(\theta) = s \left[ 1 + \sum_{n=0}^{\infty} s_{2n}(s) P_{2n}(t) \right], \qquad \text{[ZT27.8]} \quad (6)$$

where $P_n$ is a Legendre polynomial, $t = \cos\theta$, and the coefficients $s_n$, called *figure functions*, characterize the shape of the equipotential surfaces.

Equation (6) introduces a new parameter $s$ that serves as an index to the surfaces. Because the figure functions $s_n$ are not all independent, this parameter can be defined in several ways. This work uses the usual definition of $s$ as the radius of a sphere having the same volume as that enclosed by the equipotential surface,

$$s^3 = \tfrac{1}{2} \int_{-1}^{1} r(\theta)^3 dt. \qquad \text{[ZT27.6]} \quad (7)$$

Hence the average radius $s_1$, in Eq. (5), is the radius of a sphere having the same total volume as the planet. Letting $a$ and $b$ be the values of $r(\theta)$ at $\theta = \pi/2$ and $\theta = 0$, respectively, one finds

$$a/s = 1 + s_0 - \tfrac{1}{2}s_2 + \tfrac{3}{4}s_4 - \tfrac{5}{16}s_6 + \dots, \qquad \text{[ZT27.9]} \quad (8)$$

$$b/s = 1 + s_0 + s_2 + s_4 + s_6 + \dots, \qquad \text{[ZT27.10]} \quad (9)$$

where Eq. (7) relates $s_0$ to the other figure functions. Equation (8) also relates the average radius $s_1$ to the equatorial radius $a_1$ in Eq. (5).

The angular dependence of the gravitational and centrifugal potentials, $V(r, \theta)$ and $Q(r, \theta)$, can also be expressed in terms of Legendre polynomials (Eqs. 25.14 and 26.21 of ZT). Using those expressions, along with Eq. (6), and the formula for the products of Legendre polynomials (Eq. 25.8 of ZT), the total potential can be written in the form

$$U(r, \theta) = \tfrac{4}{3}\pi G\bar{\rho}s^2 \sum_{n=0}^{\infty} A_{2n}(s)P_{2n}(t),$$ 

[ZT28.3] (10)

where $\bar{\rho} = 3M_1/4\pi s_1^3$ is the average density of the planet. Since $U(r, \theta)$ is required to be constant for a given value of $s$, it follows that

$$A_{2n}(s) = 0, \ n > 0.$$ 

[ZT28.4] (11)

The solutions of these equations determine the figure functions $s_n$. The equation of hydrostatic equilibrium, Eq. (4), becomes

$$\rho^{-1}(dp/ds) = dU/ds = \tfrac{4}{3}\pi G\bar{\rho}[2sA_0 + s^2 dA_0/ds].$$ 

[ZT29.1] (12)

The solution of this equation gives the density, pressure, and other thermodynamic quantities as functions of the average radius, i.e., on the equipotential surfaces.

Complete expressions for the functions $A_0$, $A_2$, $A_4$, and $A_6$, valid to third-order in the parameter $m$, are given in Eqs. 28.7- 28.12 and 29.3-29.4 of Ref. [3]. In the interest of space, I have not reproduced those equations here. However, two points should be noted.

First, Eq. 28.12 of ZT contains a typographical error.[1] The correct equation, derived by computing Eq. (7) to third order in the parameter $m$, is

$$-s_0 = \tfrac{1}{5}s_2^2 + \tfrac{2}{105}s_2^3.$$ 

[ZT28.12] (13)

Second, ZT have dropped some terms that are of order greater than $m^3$ in deriving the expressions for $A_0$, $A_2$, $A_4$, and $A_6$. While this approximation seems justified for a third-order method, it was found to give unreasonable behavior in the figure function $s_6(s)$ for very small values of $s$. (See Figs. 5 and 6 in Sec. 7.3.) In order to investigate this problem, I made calculations that included terms of order $m^4$, but without going to a full fourth-order theory, i.e., calculation of $A_8$ and $s_8$. Inclusion of these additional terms eliminated most of the problems but had only a small ef-fect on the calculated gravitational moments. Therefore, they were not included in the calculations discussed in this report.

---

1. The error in Eq. 28.12 was not carried over into Eqs. 28.7-28.11 and 29.3-29.4, so it is obviously typographical. Errors also appear in Eqs. 30.6, 30.9, and 30.10 of Ref. [3], but those equations are not used in this work.

## 3.2 Calculated Quantities

The equations discussed above were used to compute the density, pressure, temperature, sound speed, and the three figure functions, $s_2$, $s_4$, and $s_6$, as functions of radius. Of course, these quantities can not be measured. Quantities that can be measured include the flattening, the gravitational moments, and the oscillation characteristic frequency.

The flattening measures the departure of the planetary spheroid from a sphere. It is defined in terms of the equatorial and polar radii, $a_1$ and $b_1$,

$$e = (a_1 - b_1)/a_1. \tag{14}$$

This parameter can be calculated from the figure functions at the planetary surface, using Eqs. (8) and (9).

The gravitational moments $J_n$ are related to the external gravitational potential of the planet by

$$V(r, \theta) = \frac{GM_1}{r} \left[ 1 - \sum_{n=0}^{\infty} \left( \frac{a_1}{r} \right)^{2n} J_{2n} P_{2n}(t) \right], \qquad \text{[ZT31.2]} \quad (15)$$

for $r > a_1$. In this work, the moments were computed from

$$J_n = -(s_1/a_1)^n S_n(1), \qquad \text{[ZT31.3]} \quad (16)$$

where the $S_n(1)$ are integrals, defined in Eqs. 28.10 and 28.11 of ZT, that are used in calculating the figure functions.

The oscillation characteristic frequency $\nu_0$, also called the equidistance [29], is another measurable quantity that gives insight into the validity of a planetary model. $\nu_0$ is the inverse of the time for a sound wave to propagate from the surface of the planet to the center and back again,

$$\nu_0 = \left[ 2 \int_0^{s_1} C_S^{-1} ds \right]^{-1}, \tag{17}$$

where $C_S$ is the sound speed on the planetary adiabat. Mosser, et al. [30], used Jovian seismic observations to estimate a value of $\nu_0 = 142 \pm 3$ µHz for Jupiter, although noise and other problems precluded a definitive determination. This value is significantly lower than that derived from previous planetary models, which give values in the range 152-160 µHz. However, the two-layer envelope model discussed in Sec. 7.3 is in good agreement with this measurement. No value is available for Saturn.

## 3.3 Numerical Methods

Numerical calculations of the planetary structure were carried out using a nested double iteration scheme. Given an initial guess of the figure functions and flattening parameter, Eq. (12) is integrated to determine the density and pressure as functions of radius. The integration procedure proceeds from the outer surface of the planet to the center, starting with the known mass and equatorial radius. The core region is inserted at a pressure chosen to use up all the planetary mass.

Once the density is known, Eqs. (11) are solved for the figure functions. Because these equations involve integrals of the figure functions and the density over the entire planet (Eqs. 28.7- 28.11 of ZT), an iterative procedure is used. A guess of the figure functions is made and the integrals are calculated. Equations (11) are then solved for new figure functions, and the iteration is continued until convergence is obtained. The new figure functions give a new value for the flattening.

Next, the new flattening and figure functions are used to solve Eq. (12) again, giving a new density function. The iterations are continued until the flattening and gravitational moments have converged to within 0.001%.

In all cases, the figure functions were initially set to zero to begin the iteration. The flattening could also be set to zero initially, except in cases where the core was very small. In such cases, the experimental flattening was used as an initial guess.

All calculations given in this report used a radial mesh of 200 points, the first point beginning at 0.5% of the outer radius, and the remaining points equally spaced. An additional point, at zero radius, was computed by extrapolation.

# 4. BASELINE MODELS (HOMOGENEOUS ENVELOPE)

The simplest planetary model consists of a homogeneous gas envelope with a dense core. Given the EOS model discussed in Sec. 2. and the third-order theory of figures discussed in Sec. 3., the following information is required for input to the calculation:

- the mass, equatorial radius, and rotational period,

- the abundances of He, C, N, O, and S, relative to hydrogen, and

- the temperature at 1 bar on the planetary surface.

The mass, radius, and rotational period are shown in Table 1, along with the measured flattening, the gravitational moments, $J_2$, $J_4$, and $J_6$, the equidistance (for Jupiter), and their uncertainties. All planetary models are required to match the mass, radius, and period exactly. The validity of a model, i.e., the element abundances and surface temperature, is to be judged on how well it reproduces the measured flattening, gravitational moments, and equidistance.

**Table 1: Measured Data for Jovian Planets**

| Parameter[a] | Jupiter | Saturn |
|:---:|:---:|:---:|
| Mass $M_1$ (g) | $1.8986 \times 10^{30}$ | $5.6846 \times 10^{29}$ |
| Equatorial radius $a_1$ (km) | $7.1492 \times 10^4$ | $6.0268 \times 10^4$ |
| Period $d$ (hr) | 9.9249 | 10.6549 |
| Flattening $e$ | $6.4874 \times 10^{-2}$ | $9.7962 \times 10^{-2}$ |
| $J_2$ | $1.4697 \times 10^{-2}$ ($\pm$.01 %) | $1.6332 \times 10^{-2}$ ($\pm$.06 %) |
| $J_4$ | $-5.84 \times 10^{-4}$ ($\pm$1 %) | $-9.19 \times 10^{-4}$ ($\pm$4 %) |
| $J_6$ | $3.1 \times 10^{-5}$ ($\pm$65 %) | $10.4 \times 10^{-5}$ ($\pm$48 %) |
| Equidistance $v_0$ ($\mu$Hz) | 142 ($\pm$2%) | - |

[a]All data were taken from Ref. [9], except the flattening, taken from Ref. [10], and the equidistance for Jupiter, taken from Ref. [30].

This section discusses "baseline" homogeneous-envelope models for Jupiter and Saturn that employ nominal abundances and temperatures obtained from measurements at the planetary surfaces. The results calculated using these models are in rough agreement with the data in Table 1. However, it is found that refinements and extensions must be made to bring the results into full agreement with the data. Improvements to the baseline model are considered in Secs. 5. and 7.

## 4.1  Baseline Model for Jupiter

Detailed measurements of the element abundances and temperature at the surface of Jupiter were obtained by the 1995 Galileo probe [31]-[34]. Those data were used to define the six input parameters for the baseline model, given in Table 2.

The Galileo value for the helium abundance (number of helium atoms divided by number of hydrogen atoms) is $0.0787 \pm 0.0027$ [31]. This value is 40% higher than that obtained from the older Voyager measurements [35] but 20% less than the solar abundance [36]. The deficiency of helium in the Jovian planets has led to speculation that helium is enriched in the planetary interiors. I will discuss this issue in Sec. 6.

The C, N, O, and S abundances given in Table 2 are three times the solar values. This choice is consistent with the Galileo data for C, N, and S [32][33]. By contrast, the Galileo data gave an O abundance much *smaller* than the solar value. However, that result is not considered to be characteristic of the planet as a whole, because the Galileo probe entered a very dry region of the atmosphere, a so-called "hot spot." In fact, the data showed that the oxygen abundance increases dramatically with depth and was still increasing at the time the probe ceased to function. It is generally agreed that a realistic oxygen abundance for Jupiter is roughly three times the solar value, just as for the other elements.

**Table 2: Input Parameters for Baseline Models**

| Parameter | Jupiter | Saturn |
|---|---|---|
| He/H[a] | 0.0787 | 0.026 |
| C/H[a] | $1.09 \times 10^{-3}$ | $1.45 \times 10^{-3}$ |
| N/H[a] | $3.36 \times 10^{-4}$ | $4.48 \times 10^{-4}$ |
| O/H[a] | $2.55 \times 10^{-3}$ | $3.40 \times 10^{-3}$ |
| S/H[a] | $4.86 \times 10^{-5}$ | $6.48 \times 10^{-5}$ |
| T (K) at 1 bar | 169 | 135 |

[a]Number of atoms of given element divided by number of hydrogen atoms.

The Galileo probe also measured the temperature and pressure in the Jupiter atmosphere as functions of time, to a maximum pressure of 22 bars [34]. Using these data, an EOS model for a slightly imperfect gas, the Galileo data for the composition vs. depth, and the assumptions of hydrostatic equilibrium and adiabaticity, Seiff, et al., computed a table of temperature, pressure, and density as functions of time and depth (Table 7 of Ref. [34]). The temperature and density, as functions of pressure, are shown by circles in Fig. 2.

Reference [34] gives a value of 166.1K at 1 bar. The dashed lines in Fig. 2 were computed using my EOS model, the same element abundances as in the analysis of Seiff, et al., and an initial temperature of 166K. Agreement with the temperature and density data is excellent. Note that the corrections to the rotational terms in the $H_2$ EOS, discussed in Sec. 2.1 and Appendix A, were essential in obtaining such good agreement with the data.

However, a slightly higher initial temperature, 169K, is needed to account for the difference between the element abundances used in the baseline model and those in the region entered by the Galileo probe. The dot-dashed lines in Fig. 2 were computed using the abundances in the baseline model and an initial temperature of 169K. Agreement with the temperature data is excellent. Slightly higher densities are obtained using the baseline abundances, which include higher concentrations of nitrogen and oxygen.

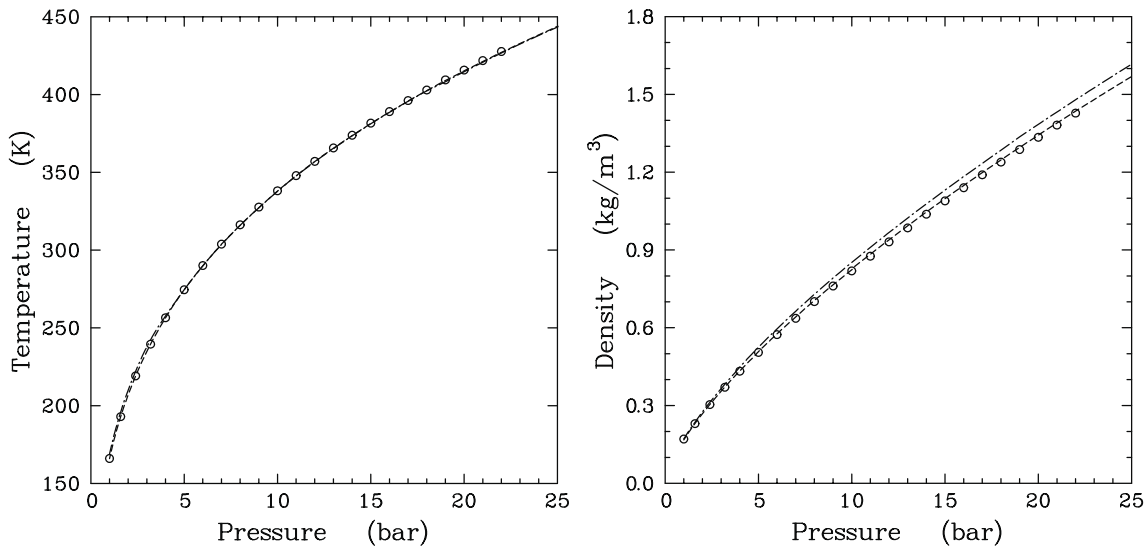

**Fig. 2.** Temperature and density vs. pressure in atmosphere of Jupiter. Circles are Galileo data from [34]. Dashed curves were computed with same abundances and 1-bar temperature as in [34], dot-dashed curves with baseline abundances and slightly higher 1-bar temperature, as explained in text.

The calculated flattening, gravitational moments and equidistance for Jupiter are given in Table 3, along with the pressure at the core boundary and the core mass and radius. The calculated moments are only in rough agreement with the measured ones; $J_2$ and $J_4$ differ from the measured values by 10% and 9%, respectively, while $J_6$ is within the error bars. The calculated flattening and equidistance are also in rough agreement with the measured one.

**Table 3: Calculated Results for Baseline Models**

| Parameter | Jupiter | Saturn |
|---|---|---|
| Flattening $e$ | $6.2711 \times 10^{-2}$ | $9.3724 \times 10^{-2}$ |
| $J_2$ | $1.3290 \times 10^{-2}$ | $1.4590 \times 10^{-2}$ |
| $J_4$ | $-5.3371 \times 10^{-4}$ | $-8.6327 \times 10^{-4}$ |
| $J_6$ | $3.3984 \times 10^{-5}$ | $8.5934 \times 10^{-5}$ |
| Equidistance $\nu_0$ (µHz) | 153.1 | 114.7 |
| Core pressure (GPa) | 4700. | 860.1 |
| Core mass $(M_C/M_E)$[a] | 25.3 | 26.6 |
| Core radius $(s_C/a_1)$[b] | 0.174 | 0.244 |

[a]Core mass $M_C$ divided by earth mass $M_E$ ($5.9742 \times 10^{27}$ g [10]).
[b]Core radius $s_C$ divided by equatorial radius $a_1$.

Clearly, the baseline model for Jupiter only provides a starting point. Significant modifications are needed to obtain predictions that are within the uncertainties in the measurements.

## 4.2 Baseline Model for Saturn

The abundance and temperature data for Saturn are more scarce and have larger uncertainties than for Jupiter. The Cassini spacecraft, which is currently making measurements of Saturn and its moons, may eventually provide better information for use in planetary modeling. In the meantime, the baseline model parameters in Table 2 were chosen using the following arguments.

The helium abundance, the most important input parameter in the model, is much more uncertain for Saturn than for Jupiter. The Voyager data for Saturn [35] gave a helium mass fraction of 0.06 ± 0.05 (relative to the total amount of helium and hydrogen). The corresponding Voyager result for Jupiter, 0.18 ± 0.04, is three times higher. By contrast, Galileo obtained a helium mass fraction of 0.238 ± 0.007 for Jupiter [31][35]. The discrepancy between the Voyager and Galileo results for Jupiter have led to questions about the validity of theVoyager data for Saturn.

The original Voyager results were obtained using a combination of radio occultation and infrared spectrometer data. Conrath and Gautier [37] attempted to redetermine the Voyager Saturn helium abundance, using only the spectral data, and obtained much higher mass fractions, in the range 0.18–0.25. Unfortunately, the approximations used in their analysis could not be tested by applying the same method to the Jupiter data. Planetary structure models also point to a larger Saturn helium abundance [6][38], but the values given in [37] appear to be *too* large, as shown in Sec. 5.2.

It is likely that theVoyager data for both Jupiter and Saturn contain systematic errors. However, it seems reasonable to accept the *ratio* of the helium abundances for the two planets, at least as a starting point. The baseline model, Table 2, takes the helium abundance in Saturn to be 1/3 that for Jupiter, as indicated by the Voyager data. The effect of varying the helium abundance is discussed in Sec. 5.2.

Current estimates of the element abundances for Saturn, relative to the solar values, are 4-6 for carbon and 2-4 for nitrogen, oxygen and sulfur being essentially unknown [32][33]. The baseline model, Table 2, assumes all four elements to be four times the solar value.

The Pioneer and Voyager spacecrafts obtained a surface temperature of 135K for Saturn, the value used in Table 2 [6]. Guillot preferred a surface temperature of 145K, arguing that a higher temperature would be inferred from the data if the helium abundance were increased over the Voyager value [6]. I have chosen to use the lower number for the baseline model, which uses a smaller helium abundance than that assumed by Guillot.

The calculated flattening, gravitational moments, equidistance, and the pressure, mass, and radius of the core for Saturn are given in Table 2. As for Jupiter, the calculated values are in rough agreement with the measured ones. $J_2$ and $J_4$ differ from the measured values by 11% and 6%, respectively, while $J_6$ is within the error bars. Once again, significant modifications of the model are needed to obtain predictions that are within the uncertainties in the measurements.

# 5. VARIATIONS USING HOMOGENEOUS ENVELOPE

The baseline models discussed in Sec. 4. give gravitational moments $J_2$ that are smaller than the measured values—10% less for Jupiter and 11% less for Saturn. These discrepancies are significantly larger than the uncertainties in the measurements, and it is desirable to determine what changes are needed to bring the models into agreement with the data.

This section discusses several ways of forcing agreement with the measured value of $J_2$ while retaining the assumption that the envelope is homogeneous with respect to the element abundances. The results show that matching $J_2$ with a homogeneous envelope model leads to a value for $J_4$ that is outside the measurement uncertainties. In Sec. 7., I will show that a two-layer envelope model *can* match all three moments, to within their uncertainties.

## 5.1 Density Scaling of Baseline Model

A straightforward way to force agreement with the measured $J_2$ is to multiply the densities along a planetary adiabat by a constant factor. It turns out that a small increase in density at constant pressure, 3-5%, is sufficient to match the measurements for both planets. But the results, shown in Table 4, are not satisfactory because the calculated values for $J_4$ are higher than the measured ones—3.5% for Jupiter and 7.2% for Saturn, well outside the uncertainties in the measurements. Also note that increasing the density on the adiabat decreases the mass and radius of the core, especially for Jupiter, but has only a small effect on the equidistance.

**Table 4: Results—Density Scaling of Baseline Models**

| Parameter | Jupiter | Saturn |
|---|---|---|
| Scale factor | 1.0340 | 1.0490 |
| Flattening $e$ | $6.4872 \times 10^{-2}$ | $9.6394 \times 10^{-2}$ |
| $J_2$ | $1.4697 \times 10^{-2}$ | $1.6332 \times 10^{-2}$ |
| $J_4$ | $-6.0440 \times 10^{-4}$ | $-9.8484 \times 10^{-4}$ |
| $J_6$ | $3.9244 \times 10^{-5}$ | $9.9476 \times 10^{-5}$ |
| Equidistance $\nu_0$ (µHz) | 149.9 | 110.8 |
| Core pressure (GPa) | 4250. | 864.0 |
| Core mass ($M_C/M_E$) | 11.6 | 21.6 |
| Core radius ($s_C/a_1$) | 0.139 | 0.230 |

## 5.2 Effect of Increasing the Element Abundances

One way to obtain an increased density along the planetary adiabat is to increase the abundances of helium and/or the heavier elements. Table 5 shows the results obtained when the helium abundance is modified to match the measured $J_2$. The calculated values are quite close to those in Table 4. In particular, the value of $J_4$ is well outside the uncertainties in the measurements.

**Table 5: Results—Baseline Models with Increased Helium Abundance**

| Parameter | Jupiter | Saturn |
|---|---|---|
| He/H | 0.10699 | 0.05697 |
| Flattening $e$ | $6.4872 \times 10^{-2}$ | $9.6395 \times 10^{-2}$ |
| $J_2$ | $1.4697 \times 10^{-2}$ | $1.6332 \times 10^{-2}$ |
| $J_4$ | $-6.0450 \times 10^{-4}$ | $-9.8430 \times 10^{-4}$ |
| $J_6$ | $3.9252 \times 10^{-5}$ | $9.9453 \times 10^{-5}$ |
| Equidistance $\nu_0$ (μHz) | 151.5 | 113.7 |
| Core pressure (GPa) | 4249. | 864.4 |
| Core mass ($M_C/M_E$) | 11.8 | 21.6 |
| Core radius ($s_C/a_1$) | 0.141 | 0.230 |

Note that the helium abundances must be increased by 36% for Jupiter and 119% for Saturn to match $J_2$. The resulting value of He/H for Jupiter is not only outside the uncertainty in the measurement, it is even larger than the solar value, 0.0977 [36]. By contrast, even with the increase, He/H for Saturn is still only 58% of the solar value. The Saturn value corresponds to a mass fraction of 0.183, close to the lower limit given by Conrath and Gautier in their re-evaluation of the Voyager data [37]. The fact that this value overestimates $J_4$ suggests that the helium mass fractions given in Ref. [37] are too large.

Alternately, one can match $J_2$ by increasing the abundances of the heavier elements while keeping the helium abundance constant. The abundances C/H, N/H, O/H, and S/H for Jupiter must be increased from 3.0 to 5.4 times the solar values, while those for Saturn must be increased from 4.0 to 6.5 times solar. These values are somewhat outside the probable uncertainties in the abundance measurements, at least for Jupiter. The results are very close to those given in Tables 4 and 5 and so are not shown. Most important, $J_4$ is still too high.

Of course, one can also match $J_2$ by increasing the values of all five abundances at the same time. It is clearly possible to match the exact value of $J_2$ for Saturn, using element abundances that are within the large uncertainties and retaining the assumption of a homogeneous gas envelope. While I have not attempted to do so, it may even be possible to do that for Jupiter, by increasing the abundances of all five elements at the same time. However, it is clear from the above results that the resulting homogeneous models would not be satisfactory for either planet, because they would overestimate $J_4$.

## 5.3 Effect of a Radiative Region in the Adiabat

Another way to increase the density along the planetary adiabat is to decrease the temperature at a given pressure. Significantly lower initial temperatures are not justified for either planet. However, Guillot, et al. [9], have argued that a radiative region, which would give a lower temperature gradient than predicted by the adiabatic model, could exist in the outer regions of both Jupiter and Saturn. For Jupiter, they predicted a radiative region extending from 0.1 to 4.2 GPa, with a temperature increase from 1300 to 2300K. For Saturn, they predicted a radiative region from 0.3 to 3.0 GPa and 1400 to 2100K.

In order to study the effects of a radiative region, the baseline adiabat was modified as follows. The original adiabat was used up to the onset temperature for the radiative region. In the radiative region, the adiabatic temperature gradient was decreased by a constant factor to obtain the same temperature offset at the end of the radiative region as obtained by Guillot, et al. At the end of the radiative region, the adiabat was continued, starting from the new density and temperature and using the adiabatic assumption.[1]

The results are shown in Table 6. In Jupiter, introduction of a radiative region brings $J_2$ closer to the measured value, but it overestimates $J_4$ by 3.3%. A radiative region has a much smaller effect on the moments for Saturn. An exact match to $J_2$ for Jupiter would require relatively small changes to the parameters, but much more drastic changes would be required to match $J_2$ for Saturn.

Whether or not a radiative region does exist in either planet may still be an open question. However, it is evident from Table 6 that the addition of a radiative region to the model does not offer a way to match *both* $J_2$ and $J_4$ in either case. Some other explanation for the planetary data must be found. Therefore, a radiative region is not considered in the remainder of this investigation.

---

1. As in Ref. [9], the radiative regions were taken to be 1300-2300K for Jupiter and 1400-2100K for Saturn, with gradient factors of 0.60 and 0.85, respectively. These parameters give different pressures for the radiative regions than those in [9]—0.10-2.6 GPa for Jupiter and 0.36-2.5 GPa for Saturn.

**Table 6: Results—Baseline Models with Radiative Region**

| Parameter | Jupiter | Saturn |
|---|---|---|
| Flattening $e$ | $6.4647 \times 10^{-2}$ | $9.4152 \times 10^{-2}$ |
| $J_2$ | $1.4554 \times 10^{-2}$ | $1.4870 \times 10^{-2}$ |
| $J_4$ | $-6.0327 \times 10^{-4}$ | $-8.8514 \times 10^{-4}$ |
| $J_6$ | $3.9501 \times 10^{-5}$ | $8.8529 \times 10^{-5}$ |
| Equidistance $\nu_0$ ($\mu$Hz) | 153.3 | 115.2 |
| Core pressure (GPa) | 4314. | 858.4 |
| Core mass ($M_C/M_E$) | 14.8 | 25.6 |
| Core radius ($s_C/a_1$) | 0.150 | 0.241 |

## 5.4 Effect of Decreased Initial Temperature

In principle, the density along the planetary adiabat could also be increased by using a lower initial 1-bar temperature. However, there is little justification for using a lower temperature for either planet. In fact, Guillot has argued for using a higher temperature for Saturn [6].

It is true that the Galileo probe measured the temperature and pressure of Jupiter in a "hot spot." However, that term is used to denote a region that is bright in the infrared because it has fewer clouds and a lower water content than surrounding areas of the surface [34]. It does not necessarily imply that the temperature in the measured region would be higher than in surrounding regions (at the same pressure). Sec. 4.1 shows that correcting the Galileo measurements to a composition containing more oxygen would actually give a slightly higher 1-bar temperature.

For completeness, however, I have made calculations for Jupiter and Saturn, using the baseline compositions but decreasing the 1-bar temperatures by 10K (159K for Jupiter and 125K for Saturn). That resulted in a 3.8% increase in $J_2$ for Jupiter and a 5.2% increase in $J_2$ for Saturn—still quite far from the measured values. The 1-bar temperatures would have to be decreased by at least 30K to bring $J_2$ into agreement with the measurements. Even then, the results for $J_4$ would be too high, just as for the other homogeneous models.

## 6. HYDROGEN-HELIUM DEMIXING

Because Jupiter and Saturn formed from the protosolar nebula, one would expect them to have roughly the same helium abundance as the sun.[1] Since the outer atmospheres of both planets have a smaller helium abundance, it has been argued that helium must be enriched in their interiors. A two-layer envelope, with a higher helium abundance in the inner region, has also been favored by planetary modelers because it helps to obtain simultaneous agreement of $J_2$ and $J_4$ with the measurements.

Demixing, or phase separation, is often cited as the mechanism for enrichment of helium in the planetary interiors. According to this theory, droplets of helium separated out of the gas mixture and were drawn into the interior by gravity, a phenomenon similar to rain.

The present study does not rule out helium enrichment in the interiors of Jupiter and Saturn. However, this section will show why demixing is not likely to be the mechanism that causes it.

Phase separation in gas-liquid mixtures is usually the result of an unfavorable energy of interaction between unlike chemical species.[2] Helium atoms are completely miscible with $H_2$ molecules at low pressures and are expected to remain so at pressures up to at least 300 GPa, where hydrogen starts to become metallic. At higher pressures, interactions between nonmetallic He atoms and metallic H atoms can lead to a large positive energy of mixing, favoring phase separation. However, the negative entropy of mixing compensates for the positive energy of mixing at sufficiently high temperatures. Hence the region of phase separation is a function of both pressure and temperature. The demix boundary is the highest temperature at which separation can occur at a given pressure.

Measurement of the He-H demix boundary is beyond the state-of-the-art and is also difficult to calculate theoretically. One of the most reasonable calculations to date is that of Pfaffenzeller, et al [40]. Their results for a 90%H:10% He mixture (solar abundance) are shown by the crosses in Fig. 3, along with the Jupiter and Saturn adiabats calculated using the baseline models. Both adiabats lie well above the demixing region, showing that phase separation cannot occur in either planet. While Saturn lies closer to the boundary than Jupiter, it should be noted that the demix boundary for a 10% He mixture is an upper limit for Jupiter and Saturn—the calculations predict even lower temperatures for smaller helium abundances.

---

1. Solar evolution models indicate that the protosolar helium abundance was less than the current value—0.0955 vs. 0.0977 [39]. This difference is small compared with the difference between the helium abundance in the planets and the sun.
2. For completeness, note that a wide disparity in molecular size can also lead to phase separation in gases [23].

None of the adiabats for the other models discussed in this report cross over into the demixing region, not even those including radiative regions, which have lower temperatures.

But, for the sake of discussion, let us suppose that the calculations of the adiabats or the demix boundary are in error and that demixing can occur in one or both of the planets. Even then, it is difficult to explain why that would lead to depletion of helium in the *outer* atmospheres, since phase separation would occur only at the high pressures in the planetary interiors. There is no mechanism for the formation of helium droplets in the outer atmospheres.[1] And, while gravity might draw helium droplets in the high-pressure *inner* regions to even greater depths, it is doubtful that helium atoms would then diffuse from the outer regions, where they are soluble in $H_2$, into the inner regions where they would (presumably) be less soluble. In fact, diffusion would be more likely to occur in exactly the opposite direction—towards the region of greater solubility.

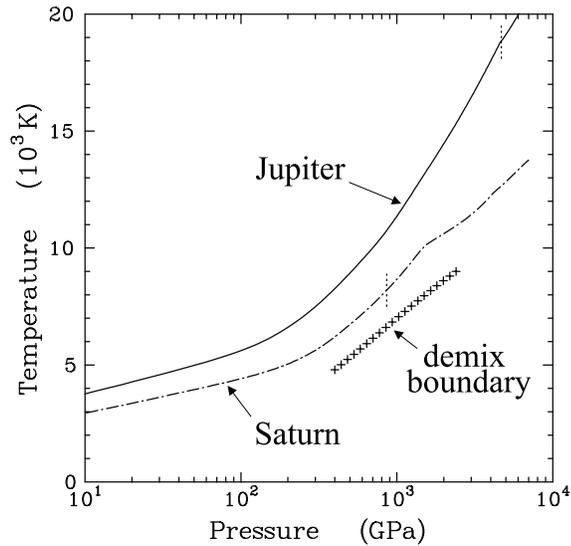

Fig. 3. Planetary adiabats compared to He-H demix boundary. Solid and dotdashed lines are baseline models for Jupiter and Saturn. Crosses show calculated demix boundary for 10% helium mixture [40]. Vertical dotted lines are core boundaries.

In Sec. 7. I will show that a two-layer envelope model, with a higher density in the inner region, will give simultaneous agreement of $J_2$ and $J_4$ with the measurements. The density increase cannot be attributed entirely to He enrichment, because that would lead to an average He/H ratio greater than the solar value—*much* greater in the case of Saturn. Therefore, at least some of the density increase must be due to enrichment of other heavy elements. He enrichment could still account for part of the density increase, but the above arguments show that demixing is not a likely mechanism for it.

---

1. Note that rain occurs on the earth when liquid water condenses out of the atmosphere. Since helium has a critical point of only 5.2 K, it cannot condense out of the atmospheres of Jupiter and Saturn, even at their low temperatures, like water does on earth.

# 7. TWO-LAYER ENVELOPE MODEL

In this section, I will show that a good match to the gravitational moments of Jupiter and Saturn can be obtained using a two-layer gas envelope, in which the inner region is denser than the outer one. This model also results in a much smaller dense core than was obtained with the homogeneous models.

I will also argue that the density change at the boundary between the layers must be due to a difference in element abundances in the two regions. If that is the case, the assumption of a sharp boundary must be regarded as an approximation to the actual planetary structure. In fact, I will show below that equally good results can be obtained if the density change is allowed to occur continuously, over a wide pressure range. Nevertheless, the two-layer model is useful because it involves the least number of input parameters and still illustrates most of the issues relevant to the structures of the two planets.

## 7.1 Two-Layer Models from Density Scaling of Baseline Models

The general features of the two-layer model can be illustrated using a simple density-scaling procedure, similar to that used in Sec. 5.1. The densities along the baseline adiabats are multiplied by a constant factor, except that the factor is applied only at pressures above that at the boundary between the two regions. The scale factor and boundary pressure are then varied until satisfactory agreement is obtained for both $J_2$ and $J_4$.

When this procedure is applied to Jupiter and Saturn, it is found that there is an upper limit to the boundary pressure. The higher the boundary pressure, the higher the scale factor must be to give agreement with $J_2$. But, as the boundary pressure and scale factor are increased, the mass of the dense core decreases. The upper limit occurs when the mass of the dense core goes to zero.

The calculations for both Jupiter and Saturn show that the best results for both $J_2$ and $J_4$ are obtained with the highest possible boundary pressure, i.e., the smallest possible dense core. Planetary formation calculations indicate that both planets should have a segregated dense core of at least 2-4 $M_E$ [12]. Table 7 shows the results obtained with a core mass of 2.5 $M_E$ and requiring an exact match to $J_2$. The values for $J_4$ are only slightly larger than the measured ones and well within the uncertainties. Only slight improvement is obtained by letting the core mass actually go to zero. The calculated values for $J_6$ are also well within the uncertainties in the measurements. The calculated equidistance for Jupiter is less than that for the baseline model, in better agreement with the measured value. This result is a consequence of the lower sound speed at the boundary of the inner and outer regions.

**Table 7: Results—Two-Layer Models Scaled from Baseline Models**

| Parameter | Jupiter | Saturn |
|---|---|---|
| Scale factor | 1.081 | 1.689 |
| Boundary pressure (GPa) | 318.90 | 227.12 |
| Flattening $e$ | $6.4885 \times 10^{-2}$ | $9.6453 \times 10^{-2}$ |
| $J_2$ | $1.4697 \times 10^{-2}$ | $1.6332 \times 10^{-2}$ |
| $J_4$ | $-5.8675 \times 10^{-4}$ | $-9.1945 \times 10^{-4}$ |
| $J_6$ | $3.7347 \times 10^{-5}$ | $9.1208 \times 10^{-5}$ |
| Equidistance $\nu_0$ ($\mu$Hz) | 148.7 | 109.1 |
| Core pressure (GPa) | 4026. | 1387. |
| Core mass ($M_C/M_E$) | 2.53 | 2.50 |
| Core radius ($s_C/a_1$) | 0.087 | 0.117 |
| Boundary mass ($M_B/M_1$)[a] | 0.760 | 0.506 |
| Boundary radius ($s_B/a_1$)[b] | 0.744 | 0.505 |

[a]Mass within inner boundary $M_B$ divided by total mass $M_1$.
[b]Radius of inner boundary $s_B$ divided by equitorial radius $a_1$.

The two-layer model predicts that *both* Jupiter and Saturn must have small segregated dense cores.[1] (The implications of this result will be discussed below.) In other respects, however, the two planets give rather different results. In Jupiter, the boundary between the inner and outer regions occurs at a pressure of 319 GPa, and the inner region encompasses 75% of the planet by mass or radius. In Saturn, the boundary pressure is somewhat lower, 227 GPa, and the inner region encompasses only 51% of the planet. Even more important, the density increase at the boundary is only 8% for Jupiter, while it is 69% for Saturn.

## 7.2  Possible Explanations for Two-Layer Model

Examination of the above results shows that the two-layer model accounts for a phenomenon related to the planetary structure, not to some error or omission in the EOS model. An EOS problem would be expected to give similar results for the density change at the inner/outer boundary in both planets. Moreover, as noted in Sec. 2.1, the $H_2/D_2$ EOS model discussed in Ref. [2] gives excellent agreement with experimental data and recent *ab initio* Hugoniot calculations for $D_2$, at pressures up to 400 GPa and temperatures comparable to those on the planetary adiabats. A density change of 1-2% might reasonably be attributed to EOS problems, but a density change of 8% cannot. And a density change of 69% is clearly far outside any reasonable estimate of EOS uncertainties.

---

1. This result is consistent with the seismic data for Jupiter, which indicate a small core [30].

Helium enrichment could also lead to an increased density in the planetary interiors. In order to investigate this possibility, I performed calculations using the same boundary pressures as in Table 7, keeping the CNOS abundances the same as in the baseline models, and choosing the He/H ratio in the inner regions to match $J_2$. For Jupiter, it was necessary to increase the He/H ratio by 72%, which corresponds to an average ratio of 0.12 for the entire planet. For Saturn, it was necessary to increase the He/H ratio by a factor of 29, which corresponds to an average ratio of 0.19. The other results were similar to those in Table 7 and so are not shown.

These results show that He enrichment cannot be the *sole* cause of the density increase in the inner region. The average He/H ratios obtained are greater than the solar abundances—20% greater for Jupiter and 90% greater for Saturn. Moreover, such a model gives much too small a mass of heavy elements. According to planetary formation calculations, the heavy element mass of both planets, including both the segregated core and the dispersed material, must be at least 12 $M_E$ [12].

It is evident that the density increase in the inner regions indicates the enrichment of elements heavier than H or He. This enrichment can also be justified on the basis of planetary formation theories, as shown below. While some enrichment of He is also possible, no mechanism to explain it is presently available, as noted in Sec. 6.

### 7.3  Two Layer Model as Result of Dispersed Planetismals

Following the work of Pollack, et al. [12], the formation of a Jovian planet can be divided into four basic phases.

I     Formation of a body large enough to begin rapid accretion.[1]

II    Rapid accretion of solid material in the solar nebula. This phase leads to a protoplanet that eventually becomes the dense, segregated core of the final planet.

III   Accretion of both solids and gases at slow rates. During this phase, gas drag, evaporation, and dynamical pressure dissolve or disperse the solid material in the gases.

IV    Runaway accretion of gases, which begins when the solid and gas accretion rates are about equal.

---

1. This phase was not considered in the simulations of Ref. [12], which began with a body the size of Mars.

Pollack, et al., presented a number of simulations, using this model with various assumptions. The exact results differed from case to case, depending upon the parameters used. However, the general picture that emerges from this model is a planet consisting of a small, dense core, 2-4 $M_E$, and a gas envelope in which the inner region has a higher concentration of dissolved or dispersed planetismals than the outer region. The two-layer envelope model presented here is an approximation to this description.

In order to proceed with the calculations, it is necessary to assume a composition for the dispersed planetismals. Pollack, et al., suggested a composition similar to that of comet Halley, i.e., 39.7% $H_2O$/30.8% rock (ferromagnesium silicates)/ 29.5% CHON (organics) by weight. The comet composition is quite close to that obtained by using the C, N, O, and Si solar abundances, assuming the additional O to be present as $H_2O$, C as $CH_4$, and N as $NH_3$, and assuming rock to have the composition $MgFeSiO_4$.[1] Hence the concentration of planetismals can be varied using a single parameter—the enrichment factor for C, N, O, S, and Si, relative to solar abundance.

The mixture calculations for the inner region of the envelope were carried out in the same way as for the outer envelope, except for the addition of Fe, $SiO_2$, and Si as chemical species. The EOS for Fe and $SiO_2$ were the same as those used in the core model (Sec. 2.4). The EOS for Si was taken from Ref. [41]. Since I do not have a suitable EOS for Mg (and its compounds) at present, I simply doubled the concentration of Fe, i.e., treated the rock as $Fe_2SiO_4$.

Table 8 shows the results obtained by varying the heavy element factor to match $J_2$, while using the same boundary pressures as in Table 7. For Jupiter, the heavy elements must be enriched by a factor of 7.473 to match $J_2$. Saturn requires an enrichment factor of 45.92. As in Table 7, the values of $J_4$ and $J_6$ are also in completely satisfactory agreement with the measured ones. The calculated equidistance for Jupiter is also in very good agreement with the measured value, a consequence of the lower sound speed at the boundary of the inner and outer regions.

The assumption of a specific composition in the inner region allows the determination of one new parameter, the total mass of heavy elements—those in the segregated core plus those dispersed throughout the inner region of the gas envelope. The total masses are comparable for the two planets—35 $M_E$ for Jupiter and 28 $M_E$ for Saturn.[2] This result appears to resolve the problem that has arisen in other planetary models that show Jupiter to have a small core and Saturn to have a large core [7][42]. In this model, both planets have small *segregated* cores. The additional solid material that is expected from planetary formation theories is dispersed in the inner region of the gas envelopes and of comparable mass in both planets.

---

1. The combined abundances of Mg and Fe are 1.974 times the abundance for Si [36].
2. Most of the calculations in Ref. [12] predicted the total mass of heavy elements to be about half of that obtained here. However, the initial surface density of planetismals—a model parameter—was chosen to give masses in agreement with those predicted by previous planetary models. A 50% increase in that parameter was shown to give results close to those predicted by this model.

**Table 8: Results—Two-Layer Models Using Dissolved Planetismals**

| Parameter | Jupiter | Saturn |
|---|---|---|
| Heavy element enrichment[a] | 7.473 | 45.92 |
| Boundary pressure (GPa) | 318.90 | 227.12 |
| Flattening $e$ | $6.4885 \times 10^{-2}$ | $9.6452 \times 10^{-2}$ |
| $J_2$ | $1.4697 \times 10^{-2}$ | $1.6332 \times 10^{-2}$ |
| $J_4$ | $-5.8706 \times 10^{-4}$ | $-9.1927 \times 10^{-4}$ |
| $J_6$ | $3.7367 \times 10^{-5}$ | $9.1212 \times 10^{-5}$ |
| Equidistance $\nu_0$ ($\mu$Hz) | 145.8 | 101.8 |
| Core pressure (GPa) | 4038. | 1356. |
| Core mass ($M_C/M_E$) | 3.03 | 2.98 |
| Core radius ($s_C/a_1$) | 0.092 | 0.124 |
| Boundary mass ($M_B/M_A$) | 0.760 | 0.506 |
| Boundary radius ($s_B/a_1$) | 0.744 | 0.505 |
| Heavy element mass ($M_Z/M_E$)[b] | 34.94 | 28.32 |

[a]Ratio of heavy element abundance ($CNOS + Fe_2SiO_4$) to solar abundance.
[b]Mass of heavy elements in core and inner envelope divided by earth mass.

Figure 4 compares the planetary adiabats for the two-layer and homogeneous envelope (baseline) models of Jupiter and Saturn. The two-layer model differs only slightly from the baseline model for Jupiter (an 8% shift in density above 319 GPa), although the differences are significant when used to compute the planetary moments. The differences are much larger for Saturn; the large density increase above 227 GPa suggests that the inner region can be regarded almost as much a part of the core as a part of the gas envelope.

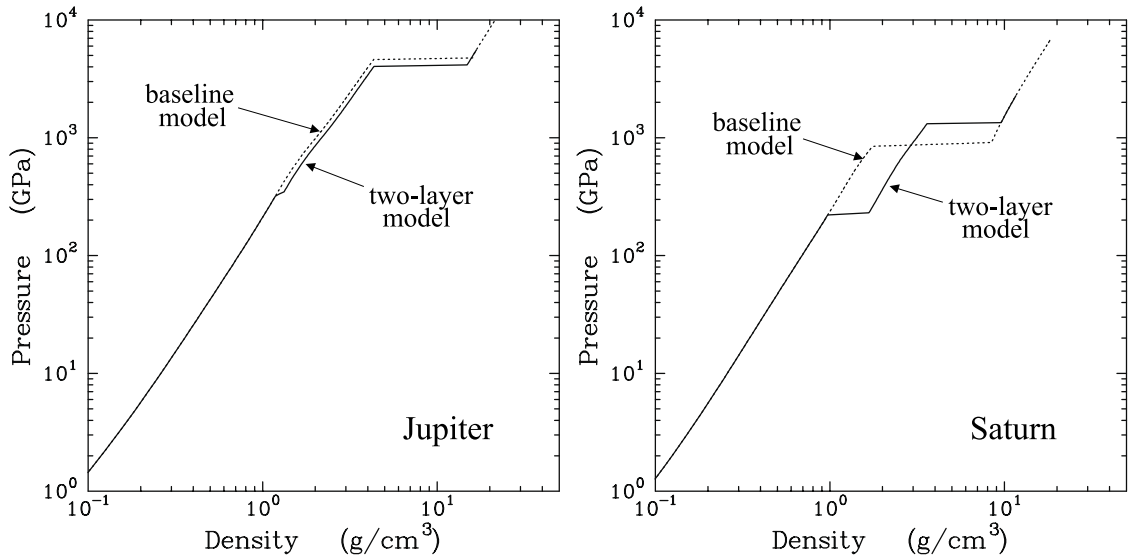

**Fig. 4. Pressure vs. density on planetary adiabats for Jupiter and Saturn. Solid lines are calculated using two-layer envelope models, dotted lines with homogeneous envelope baseline models.**

Figures 5 and 6 show the density, pressure, temperature, and figure functions vs. average radius for Jupiter and Saturn, computed using the two-layer envelope models. The boundaries of the core and inner envelope regions are also shown as vertical dotted lines. As in Fig. 4, the jump in density at the boundary of the inner envelope is much more pronounced for Saturn than for Jupiter.

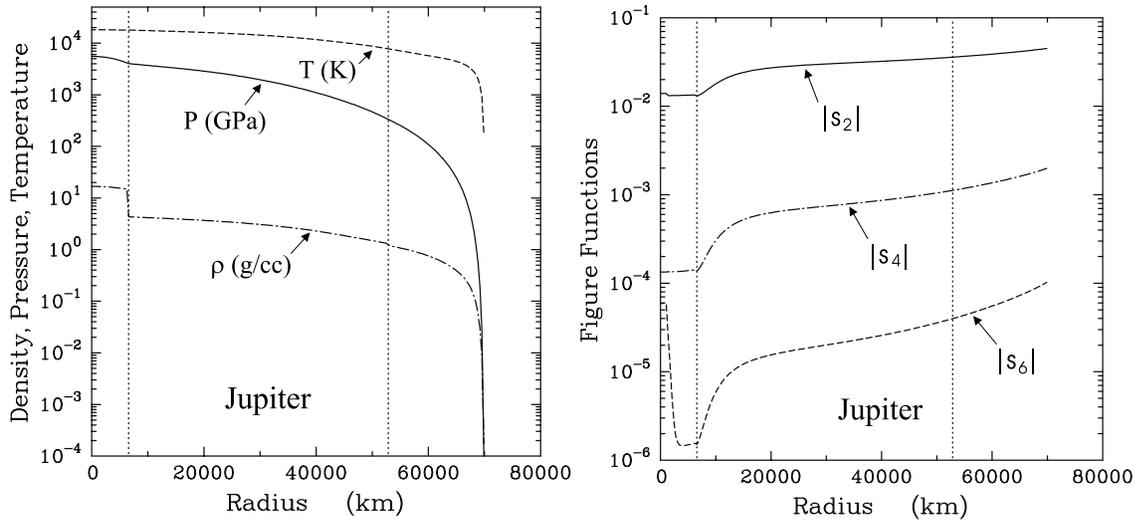

**Fig. 5. Structural parameters vs. average radius for Jupiter. Density, pressure, temperature and figure functions were computed using the two-layer envelope model. Core and inner envelope boundaries are shown by vertical dotted lines.**

Figure function $S_6$ shows a sharp rise within the core region, near zero radius, in both planets. This result is anomalous and appears to be due to truncation of the perturbation series. As noted in Sec. 3.1, the addition of higher-order terms to the equations for the figure functions eliminates most of this effect but does not alter the chief predictions of the model.

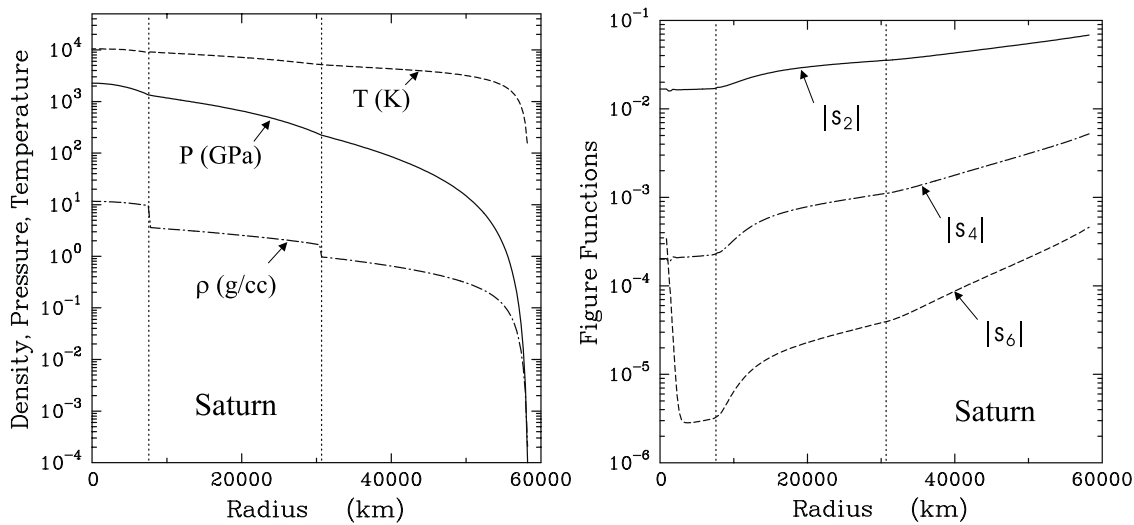

**Fig. 6. Structural parameters vs. average radius for Saturn. Density, pressure, temperature, and figure functions were computed using the two-layer envelope model. Core and inner envelope boundaries are shown by vertical dotted lines.**

Two other sets of calculations were made to investigate the importance of the planetismal composition in the results.

- In one set, the planetismals were assumed to consist entirely of CNOS, having the same relative abundances as in the outer region. In order to match $J_2$, the solar abundances had to be multiplied by a factor of 10 for Jupiter and 63 for Saturn. All other results were nearly identical to those in Table 8.
- In the second set, the CNOS was taken to be the same as in the outer region, and the extra planetismal material was assumed to consist entirely of $Fe_2SiO_4$. In order to match $J_2$, the solar abundance of $Fe_2SiO_4$ had to be multiplied by a factor of 18 for Jupiter and 140 for Saturn. Once again, the other results were nearly identical to those in Table 8.

These sensitivity tests show that an exact knowledge of the planetismal composition is not crucial to the model calculations. While that is fortunate in one sense, a way of determining the composition might actually be useful to planetary origin theories.

## 7.4 Two-Layer Model with Transition Region

As noted above, planetary formation models support the idea that the inner region of the gas envelope should have a higher concentration of dissolved planetismals, and hence a higher density. However, the existence of a sharp density discontinuity at the boundary of the inner and outer layers appears unrealistic and artificial.

In order to demonstrate that the central conclusions of this study do not require the existence of a sharp boundary, calculations were also made in which a transition region was inserted between the inner and outer layers. This region was constructed by removing a section of the adiabat and replacing it with an interpolated curve.[1] The upper pressure of the transition region was arbitrarily chosen to be roughly 2.3 times that at the inner/outer boundary, and the lower pressure was then chosen to match the measured value of $J_2$.

Table 9 shows the calculations for both planets. The results are quite close to those in Table 8. The gravitational moments are all in satisfactory agreement with the measured values. The equidistance for Jupiter is not as good as that obtained with the sharp boundary. However, this parameter could probably be brought into line by using a slightly different interpolation scheme in the transition region.

**Table 9: Results—Two-Layer Models with Transition Regions**

| Parameter | Jupiter | Saturn |
|---|---|---|
| Pressure range (GPa) | 88.268-745.05 | 103.90-543.25 |
| Flattening $e$ | $6.4884 \times 10^{-2}$ | $9.6450 \times 10^{-2}$ |
| $J_2$ | $1.4697 \times 10^{-2}$ | $1.6332 \times 10^{-2}$ |
| $J_4$ | $-5.8784 \times 10^{-4}$ | $-9.2189 \times 10^{-4}$ |
| $J_6$ | $3.7447 \times 10^{-5}$ | $9.1358 \times 10^{-5}$ |
| Equidistance $\nu_0$ ($\mu$Hz) | 149.9 | 104.6 |
| Core pressure (GPa) | 4053. | 1359. |
| Core mass ($M_C/M_E$) | 3.31 | 4.03 |
| Core radius ($s_C/a_1$) | 0.096 | 0.136 |

---

1. The interpolation was constructed using a standard option in the Panda code. This method gives a roughly linear relation between the logarithmns of pressure and density. The sound speed was computed by numerical differentiation.

# 8. CONCLUSIONS

This report shows that good agreement with the gravitational moments for Jupiter and Saturn can be obtained using a two-layer gas envelope, the interior of which is highly enriched in heavy elements, together with a small segregated core. Both planets have a core mass of ~$3M_E$, a small value compared to previous planetary models. However, the total mass of heavy elements in the core and inner envelope is ~$30M_E$, large enough for an accretion model of planetary formation. The new model also gives good agreement with the Jupiter oscillation frequency, and the prediction of a small central core is consistent with the Jupiter seismic data [30].

The existence of a two-layer gas envelope can be justified using the model of planetary origin of Pollack, et al. [12]. In the early stages of formation, accretion of solids from the protosolar nebula created the small segregated core of the planet. Both solids and gases were accreted in the next stage, creating the inner region of the gas envelope with a high abundance of heavy elements. Accretion of gases predominated in the final stage, creating the outer region of the envelope with a lower abundance of heavy elements.

The two-layer envelope is clearly a simplification of the structure, because the heavy element abundances would be expected to vary continuously as a function of radius. The development of a more realistic model for the distribution of heavy elements will probably require consideration of the formation and evolution processes. The present work could also be used to make a number of improvements to the planetary formation calculations.

The present investigation does not explain why the atmospheres of Jupiter and Saturn are deficient in helium, when compared with the sun. The standard argument, that hydrogen-helium phase separation leads to an enrichment of He in the interior, is shown to be unlikely. It is also shown that He enrichment cannot be the sole cause of a high-density inner region of the gas envelope. It remains to be shown whether or not another mechanism for He enrichment exists. If not, it is necessary to explain why the protosolar nebula did not have a uniform helium abundance when the giant planets were formed.

# ACKNOWLEGEMENTS

Many, *many* thanks to the reference staff of the Lynchburg, VA library, for procuring most of the papers and books used in this work; I could not have done this project without their help. As always, I thank my wife Donna for her prayers, her encouragement, and for acting as a sounding board for my ideas. Finally, to the extent this work has any value, all the credit belongs to the Lord.

# APPENDIX A

# Rotational Corrections to the $H_2$ EOS

An exact calculation of the rotational contributions to the $H_2$ EOS requires treatment of the coupling between the rotational degrees of freedom and the nuclear spins. This coupling arises because hydrogen nuclei have a spin of 1/2 and are fermions. Therefore, the product of the rotational and nuclear wavefunctions for the $H_2$ molecule must be antisymmetric with respect to exchange of the nuclei.

The $H_2$ EOS model of Ref. [2] did not account for this effect, which is important at temperatures below about 500K. This appendix explains how the model was corrected to account for the nuclear spin effects. The treatment used here is based on the discussion given by McQuarrie [43].

There are four possible pairing of the nuclear spins in the $H_2$ molecule, resulting in one antisymmetric spin function and three symmetric spin functions. In para-$H_2$, the spins have opposite directions, the spin wavefunction is antisymmetric, and only even values of the rotational quantum number $j$ are allowed. In the rigid rotator approximation, the partition function for a molecule of para-$H_2$ is

$$q_p = \sum_{j \text{ even}} (2j + 1) \exp[-\beta B_e j(j + 1)], \tag{A.1}$$

where the notation is the same as in [2]. In ortho-$H_2$, the spins are parallel, the spin wavefunction is symmetric, and only odd values of the rotational quantum number $j$ are allowed. The partition function for a molecule of ortho-$H_2$ is

$$q_o = 3 \sum_{j \text{ odd}} (2j + 1) \exp[-\beta B_e j(j + 1)], \tag{A.2}$$

where the factor of three comes from the three symmetric spin functions.

Now consider a mixture of $N_p$ para molecules and $N_o$ ortho molecules, where the total number of molecules is $N = N_p + N_o$. The rotational partition function for the mixture is

$$Q_{rot} = \frac{N!}{N_p! N_o!} \frac{q_p^{N_p} q_o^{N_o}}{4^N}. \tag{A.3}$$

Here the term involving the factorials is the usual statistical weight for a mixture. $Q_{rot}$ is divided by a factor of $4^N$ to correct for the statistical weight due to nuclear spin states. (If this correction were not included, a nuclear spin factor would have to be included in the EOS for atomic hydrogen and all the other hydrogen-containing compounds in the chemical equilibrium model of Sec. 2.3.)

Next define $f = N_o/N$ to be the fraction of molecules in the ortho state. Using Stirling's approximation for factorials, the rotational contribution to the Helmholtz free energy is

$$A_{rot} = -kT \ln Q_{rot} = -NkT[(1-f)\ln q_p + f\ln q_o - (1-f)\ln(1-f) - f\ln f - \ln 4] . \text{(A.4)}$$

The equilibrium concentration is obtained by minimizing the free energy with respect to $f$ (at constant temperature). The result is

$$f = 1/(1 + q_p/q_o) . \tag{A.5}$$

The rotational contributions to the internal energy and entropy are given by

$$E_{rot} = -T^2 \left[ \frac{\partial(A_{rot}/T)}{\partial T} \right]_f = NkT \left[ (1-f)T\frac{\partial \ln q_p}{\partial T} + fT\frac{\partial \ln q_o}{\partial T} \right] , \tag{A.6}$$

$$S_{rot} = (E_{rot} - A_{rot})/T . \tag{A.7}$$

There is no contribution to the pressure, because the free energy is independent of density.

The corrections to the rotational energy and entropy of $H_2$ were computed by subtracting the values obtained from the approximate model of Ref. [2] from the results obtained using Eqs. (A.6) and (A.7). Figure 7 shows the correction terms, $\Delta E_{rot}$ and $\Delta S_{rot}$, as functions of temperature from 0 to 1000K. [1]

The corrections were put into tabular form and added to the $H_2$ EOS of [2] using a standard option in the PANDA code.

---

1. The procedure used here leads to a negative entropy at low temperatures. This result is due to the fact that the rotational partition function has been divided by the nuclear spin factor, $4^N$, as discussed above. The entropy would be positive everywhere if the nuclear spin factor were included, but that would necessitate corrections to the other hydrogen compounds as well.

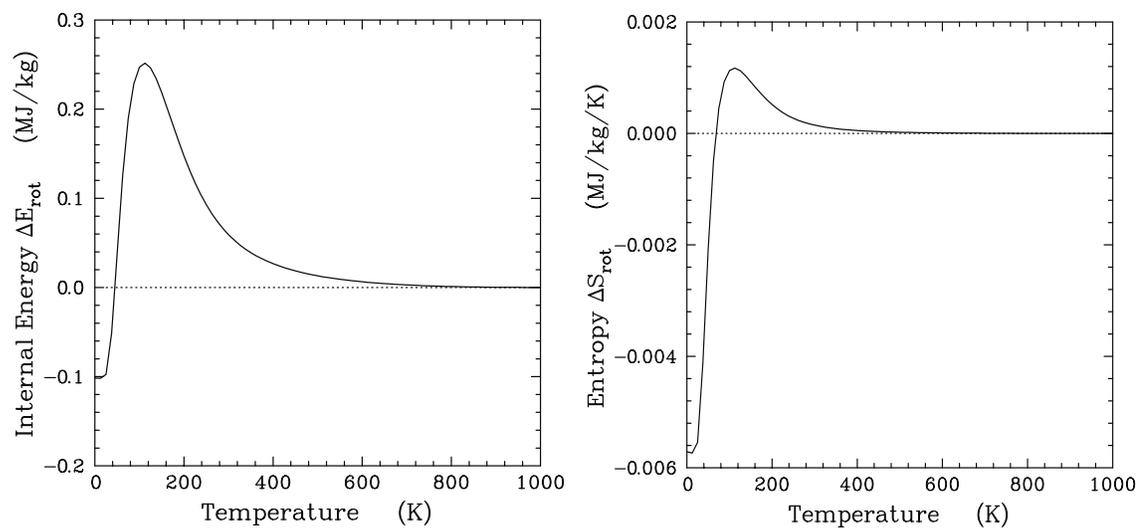

**Fig. 7.** **Nuclear spin corrections to rotational energy and entropy for molecular hydrogen.**